# TiO$_2$ multi-leg nanotubes for Surface-enhanced Raman scattering

Harini S[1], Garima Gupta[2], Somnath C. Roy[2], Rambabu Yalavarthi[1,±,*]

[1]Department of Physics, School of Advanced Sciences, Vellore Institute of Technology Vellore, Tiruvalam road, Katpadi-632014, Tamilnadu, India.

[±] Current address: Nanoinstitute of Munich, Faculty of Physics, Ludwig-Maximilians-Universität München, München-80539, Germany.

[2] Semiconducting Oxide Materials, Nanostructures and Tailored Heterojunctions (SOMNaTH) Lab, Department of Physics, Indian Institute of Technology Madras, Chennai-600036, Tamilnadu, India.

[*]Corresponding author: rambabu.y@vit.ac.in; r.yalavarthi@physik.uni-muenchen.de

**Abstract**

In the recent past, significant research efforts have been put forth to fabricate low-cost noble metal-free substrates for surface-enhanced Raman spectroscopy (SERS) applications. Here we propose semiconducting $TiO_2$ multi-leg nanotubes ($TiO_2$ MLNTs, with and without the gold nanoparticle coating) as SERS substrates. $TiO_2$ MLNTs show unique multi-leg morphology compared to the conventional non-multi-leg tubes and possess better light-harvesting properties. $TiO_2$ MLNTs are fabricated with a simple and versatile single-step electrochemical anodization method. Remarkable high SERS sensitivity is observed towards the detection of Methylene blue (MB), up to nM concentration (E.F. ~$10^4$). The same is attributed to the resonantly matched photonic absorption edge of $TiO_2$ MLNTs with the wavelength of incident laser probe light. On the other hand, gold nanoparticle-coated $TiO_2$ MLNTs demonstrated further enhancement in SERS sensitivity (E.F. ~$10^5$, for nM of MB) facilitated by the synergy that exists between the plasmonic modes (LSPRs) of Au and the photonic absorption mode of $TiO_2$ MLNTs.

**TOC graphics.**

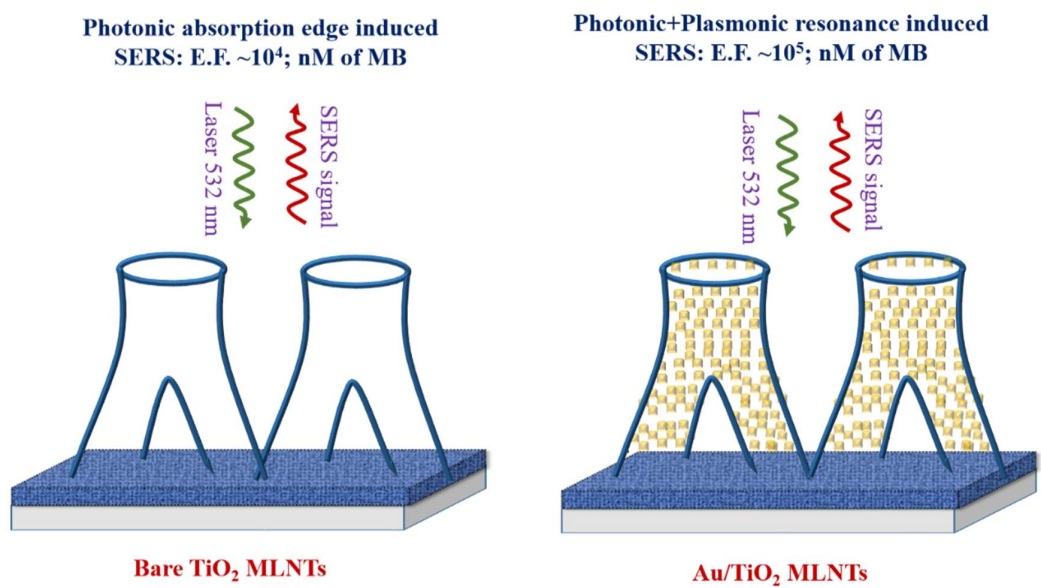

Surface-enhanced Raman scattering (SERS) is a powerful non-destructive technique that allows highly sensitive detection of trace amounts of chemical and biological analytes.[1,2] SERS has found its applications in material sciences,[3] chemical analysis,[4] biomedical research,[5] forensics,[6] etc. Recent developments have led to molecular imaging,[7] photo-electrochemical characterizations,[8,9] and even single-molecule detection.[10,11]

SERS relies on the interaction between incident light and the nanostructured substrates of noble metals, for example, Au, and Ag.[1] The SERS sensitivity of molecules adsorbed on a noble metal substrate is attributed to the electromagnetic (EM) enhancement associated with a charge transfer process (chemical enhancement) at the EM "hot spots".[12] EM enhancement is proportional to the electric field (hot spots) generated by the resonantly excited free electrons of the metal surface upon the interaction with light.[1] In the vicinity of these hot spots, the plasmon-induced near fields effectively couple to the vibrational modes of adsorbed molecules and thus amplify the Raman signal cross-section several times.[13] The magnitude of the electric field intensity depends on the size, shape, gap, and composition of the metal nanostructure.[9,14,15] Therefore, to achieve better SERS sensitivity with higher enhancements in a reproducible manner, it is indeed necessary to fabricate the tailored nanostructures of metals.[13,16] However, due to the high cost, low abundance, biocompatibility, and re-usability, the use of noble metals is limited in the fabrication process of SERS substrates.[13]

In this context, semiconductor nanostructures have been proposed to minimize the metal content in the fabrication of SERS substrates.[17,18] These nanostructures either themselves act as SERS substrates or provide structural support for the anchoring of noble metals.[17,19,20] As such, the metal content can be minimized significantly. The SERS intensity of pristine semiconductors (e.g. $TiO_2$, ZnO) is generally attributed to the enhanced light-matter interaction facilitated by the photonic band gap.[17,19–21] Qi et al reported noble metal-free $TiO_2$ photonic microarray SERS substrates. The SERS activity of such photonic band nanostructured substrates (PBNs) is mainly attributed to the resonantly matched photonic absorption edge with the incident wavelength of laser probe light.[17] Thus, the main criterion for SERS enhancement in PBNs is the tuning of the photonic absorption edge in correlation with the wavelength of incident laser probe light.[20] The photonic band edge in semiconductors can be altered by modifying the nanostructure's shape, size, and periodicity.[22] However, the enhancement factors of PBNs are generally low when compared to the noble metal SERS substrates.[17] To address this problem, a strategy for making photonic-plasmonic nanocomposites (noble metal nanostructure anchored PBNs) has been proposed. The LSPRs associated with the noble metal nanostructures synergistically couple to the photonic band edge and thus greatly enhance the SERS intensity. Till now, only a few studies focused on the fabrication of SERS substrates of LSPRs-coupled PBNs. For example, Franz et al reported Au nanoparticle decorated porous silica photonic substrates for enhanced SERS.[19] In another report by Wang et al, photonic-plasmonic microspheres (PPMs) composed of $SiO_2$ nanospheres decorated with gold nanoparticles have been studied for enhanced SERS.[20] Zheng et al explored the Au nanocrystal-incorporated bio-inspired photonic structures.[23] Ben-Jaber et al designed several Au-semiconductor systems for studying with photo-induced SERS (PIERS) the oxygen vacancies dynamics and their impact on photocatalysis.[24–26] However, the fabrication procedures of such tailored nanostructures are complex and require expensive equipment, e.g., lithography. Other fabrication procedures such as template-assisted methods and self-assembly of nanoparticles over solid surfaces are time-consuming and it is difficult to scale up into large dimensions.[27,28]

Here, we propose $TiO_2$ multi-leg nanotubes ($TiO_2$ MLNTs), as well as gold nanoparticle-coated $TiO_2$ MLNTs (Au/$TiO_2$ MLNTs) as SERS substrates. The method of fabrication process of these substrates is simple and achieved via single-step electrochemical anodization, followed by an Au layer (thickness, 10 to 30 nm) sputtered over the nanotubes and subsequent annealing in air ambient.[29,30] The substrates demonstrated excellent SERS activity towards the detection of Methylene blue (MB). We discuss the role of multi-leg morphology towards the detection of MB (SERS; up to nM concentration) as well as the synergy that exists between photonic and plasmonic modes of Au/TiO$_2$ MLNTs and their influence in enhancing the SERS sensitivity in connection with the light absorption spectra.

TiO$_2$ MLNTs were synthesized by the electrochemical anodization method.[29,30] Briefly, Titanium foil (20 mm x 20 mm x 0.25 mm; 99.7% purity, Sigma-Aldrich) was used as an anode, and a Pt foil of a similar dimension was used as a cathode. Initially, Ti foils were cleaned in de-ionized water, ethanol, and acetone for 5 minutes each with an ultrasonic cleaner, and dried in ambient air. A mixture of 96 ml di-ethylene glycol (DEG, Merck), 4 ml of de-ionized water, and 0.6 wt% of ammonium bi-fluoride salt (NH$_4$HF$_2$, Merck) was stirred in a 100 ml beaker using a magnetic stirrer for 30 minutes and used as the electrolyte solution. The anodization was carried out for 2 hours with a constant voltage of 60 V at room temperature of about 30 °C. After anodization, the samples were rinsed thoroughly with isopropyl alcohol and de-ionized water and kept at room temperature for 1 hour to dry. Subsequently, a thin film of gold was deposited on the as-obtained TiO$_2$ MLNTs by sputtering technique. The deposition was carried out for a period of 30 to 90 seconds to vary the thickness of the gold film. Eventually, the gold layer-coated TiO$_2$ MLNTs were annealed at 450 °C in a tubular furnace for 2 hours with the ramp setting of 1 °C/min. During the process of annealing, the gold film deposited over TiO$_2$ nanotubes turned into gold nanoparticles due to the de-wetting phenomenon and as a result, Au nanoparticle-coated TiO$_2$ MLNTs were formed. A probe analyte, MB was used to investigate the SERS enhancement. Initially, MB was deposited by immersing different substrates (Bare TiO$_2$, and 30 s, 60 s and 90 s Au deposited TiO$_2$ MLNTs) in 20 ml of varying molar concentrations of MB (nM to mM) for 24 hours.

X-ray diffraction (XRD) analysis was executed in a 2θ range of 20° to 60° by Bruker D8 Advance X-ray diffractometer equipped with a 2.2 KW Cu Kα anode source. A scanning electron microscope (SEM, FEI Inspect 550) was used to analyze the surface morphology, and Image-J software was used to analyze the size of nanostructures. Similarly, energy dispersive spectroscopy (EDS) measurements were performed to determine the elemental composition of the samples. UV-Vis diffuse reflectance measurements (DRS) were performed using Perkin Elmer Lambda 950 spectrophotometer equipped with an integrating sphere. Raman measurements were performed using a Renishaw Raman microscope to analyze the SERS enhancement. All spectra were recorded with an acquisition time of 10s and measured on at least three different spots of the sample. The laser power (2 mW) and beam spot are kept constant throughout all measurements. Origin software was used for baseline correction and peak analysis.

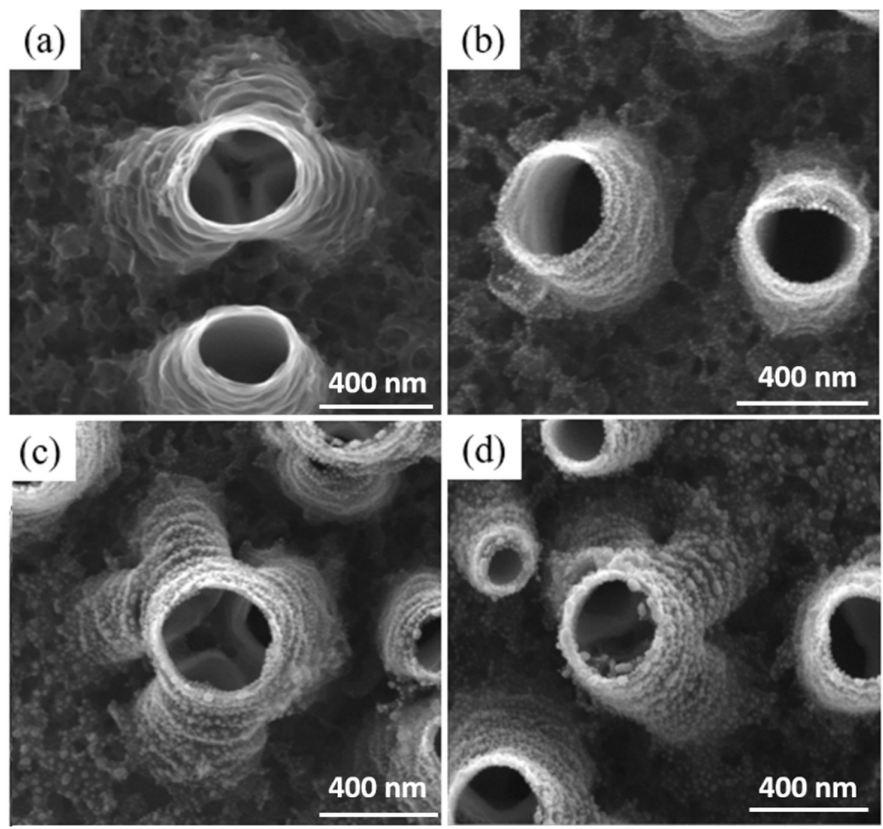

**Figure 1**: FE-SEM images of bare and Au nanoparticle-coated TiO$_2$ MLNTs. (a) bare TiO$_2$ MLNTs. (b) Au/TiO$_2$ MLNTs; Au-30 s. (c) Au/TiO$_2$ MLNTs; Au-60 s. (d) Au/TiO$_2$ MLNTs; Au-90 s.

Figure 1 shows the FE-SEM images of bare and Au nanoparticle-coated TiO$_2$ MLNTs. Figure 1(a) demonstrates the branched morphology of a multi-leg nanotube structure with the combination of three or more tubes fused at the top, with a pore diameter ranging between 120±5 nm to 330±6 nm, and the average diameter is ~200 nm. Such a multi-leg morphology offers a change in effective refractive index ($n_{eff}$) from top to bottom.[29] The gradient in $n_{eff}$ along the tube length allows the light to scatter multiple times within the MLNTs, thus significantly enhancing the light-matter interaction. As such, TiO$_2$ MLNTs resemble the PBNs with a photonic absorption edge. The surface morphology of Au deposited (for 30 s, 60 s, and 90 s) TiO$_2$ MLNTs by sputtering technique followed by annealing have been presented in Figure 1 (b), (c), and (d) respectively. The morphological observations reveal that the spherical Au nanoparticles are distributed homogenously on the top and side walls of the nanotubes. The size of Au nanoparticles varies from 10±2 nm to 45±3 nm with the change in deposition time from 30 to 90 s.

X-ray diffraction patterns shown in Figure 2a demonstrate the phase composition and crystal structure of TiO$_2$ and Au/TiO$_2$ MLNTs. The characteristic peak at a 2θ value of 25.28° corresponding to the (110) plane indicates the anatase crystal phase of TiO$_2$ in both samples (JCPDS: 87-0920).[30] In addition, the predominant peaks appeared at 2θ values of 35.24°, 38.58°, 40.30°, and 53.11° correspond to (100), (002), (101), and (102) crystal planes represent the Titanium substrate (JCPDS: 89-2762). XRD peak corresponding to metallic Au nanoparticles at a 2θ value of 38.10° is overlapped with the peak corresponding to the Titanium at the same position.[14] This can be understood by observing the difference in intensities of the peak that appeared at 38.1° for TiO$_2$ and Au/TiO$_2$ MLNTs.

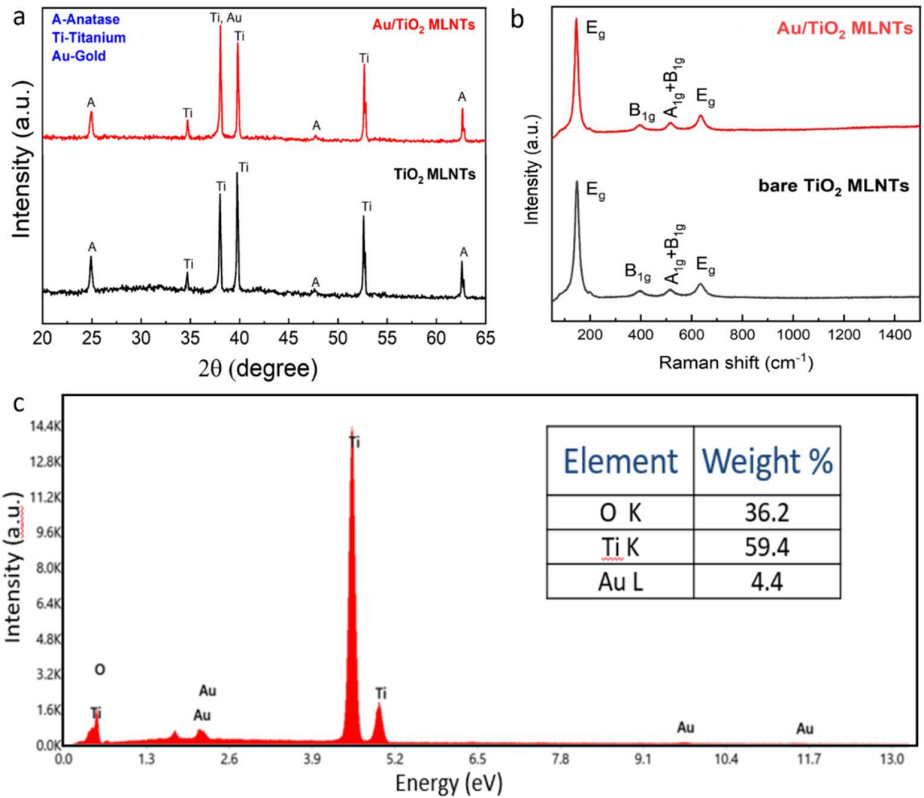

**Figure 2**: (a) XRD patterns of TiO$_2$ MLNTs with and without the Au nanoparticles coating (Au-60 s). (b) Raman spectra of TiO$_2$ MLNTs with and without the Au nanoparticles coating (Au-60 s). (c) EDS spectrum of Au /TiO$_2$ MLNTs (Au-90 s).

Figure 2b shows the Raman spectra of TiO$_2$ and Au/TiO$_2$ MLNTs, which are recorded without the MB deposition over the MLNTs. The spectra confirm the anatase phase of TiO$_2$ MLNTs with the Raman vibrational modes corresponding to $E_g$ at 143 cm$^{-1}$, $B_{1g}$ at 513 cm$^{-1}$, and $A_{1g}+B_{1g}$ at 638 cm$^{-1}$.[30] Figure 2c, EDS spectra shows the elemental composition of 90 s Au sputtered TiO$_2$ MLNTs. The percentage (Wt%) of elements, Titanium (Ti) is about 59.4 %, oxygen (O) is 36.2 % and Au is 4.4 %, which reveals the reduced content of expensive noble metals (Au) in the fabrication process of SERS substrates.

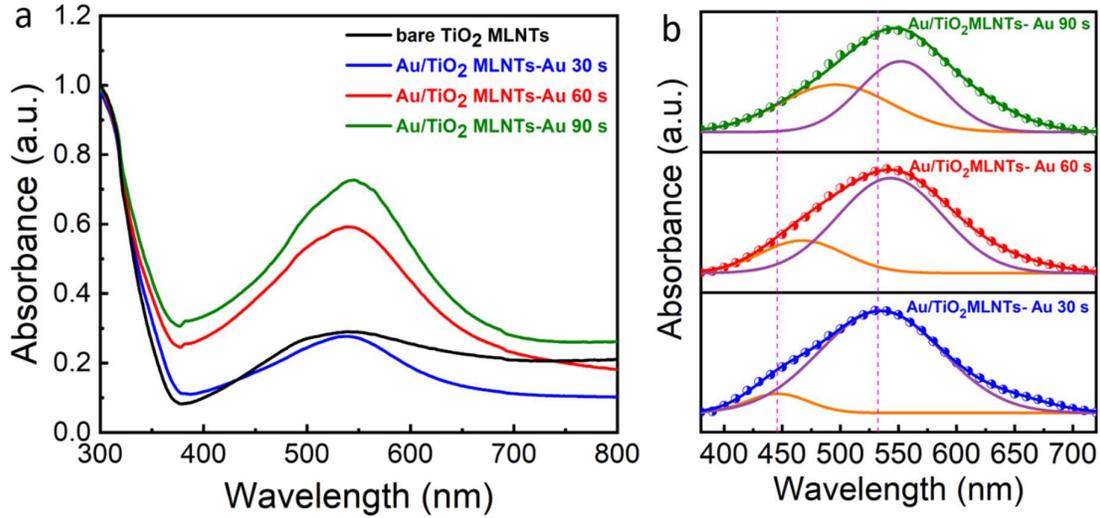

**Figure 3**: (a) DRS spectra (absorbance) of TiO$_2$ and Au/TiO$_2$ MLNTs. (b) Lorentzian fitting of absorption peak (400-700 nm), corresponding to Au/TiO$_2$ MLNTs.

Figure 3a shows the absorbance spectra (1-R) of TiO$_2$ MLNTs with and without the Au nanoparticle coating obtained from the DRS measurements. The absorption edge at ~390 nm represents the electronic band gap of bare TiO$_2$ MLNTs, which is ~ 3.16 eV. In addition, the spectra also show broad absorption corresponding to the wavelength range 400-700 nm. Defects (oxygen vacancies) could be the reason for such an absorption. However, previous studies ruled out the possible existence of defects in TiO$_2$ MLNTs.[31] On the other hand, it has been shown that the coefficient of specular reflection (R) reduced by a factor of 102 in comparison to the non-multi-leg tubes due to the change in n$_{eff}$ (top section-1.06; middle section-1.14; and bottom section-2.49 of lengths 1.2, 1.6 and 0.2 $\mu$m respectively) along the tube length.[29] Therefore, we attribute such a broad absorption to the characteristic gradient refractive index profile of TiO$_2$ MLNTs facilitated by the multiple reflections of incident light from the Ti substrate.[29] As such, the gradient in n$_{eff}$ slows down the velocity of incident light within the layered morphology and generates a photonic absorption edge peaking at 535 nm and thus significantly enhances the light-matter interaction.[17,21] Similarly, a sharp peak at ~ 535-550 nm is measured for Au/TiO$_2$ MLNTs (Au-30 to 90 s). A linear increment in intensity and a red shift in peak position is observed with the change in Au deposition time, which we attributed to the variation in Au nanoparticle size upon an increase in deposition time. To better understand, we have de-convoluted the peak using Lorentzian fitting and spectra are shown in Figure 3b. The fitting peak of wavelength range 440-475 nm corresponding to the energy 2.6-2.8 eV, is attributed to the characteristic inter-band, *d-sp* transitions of Au, and is thus responsible for the LSPRs associated with the Au nanoparticles.[14,16] In general, for pristine Au nanostructures, the intense local electric fields corresponding to LSPRs generate so-called *"hot spots"* in the vicinity of the probe molecules and thus enhance the Raman absorption cross-section by many folds.[14,16] However, apart from the LSPR peak, there is another fitting peak in the range 535 to 550 nm corresponding to the energy ~ 2.5 to 2.3 eV. This is in line with the photonic absorption edge (~535) nm of TiO$_2$ MLNTs as discussed above. Therefore the coupling between the photonic mode and LSPR peak generates a new hybrid cavity mode, which is the absorption peak observed for Au/TiO$_2$ MLNTs and is responsible for the SERS enhancement.

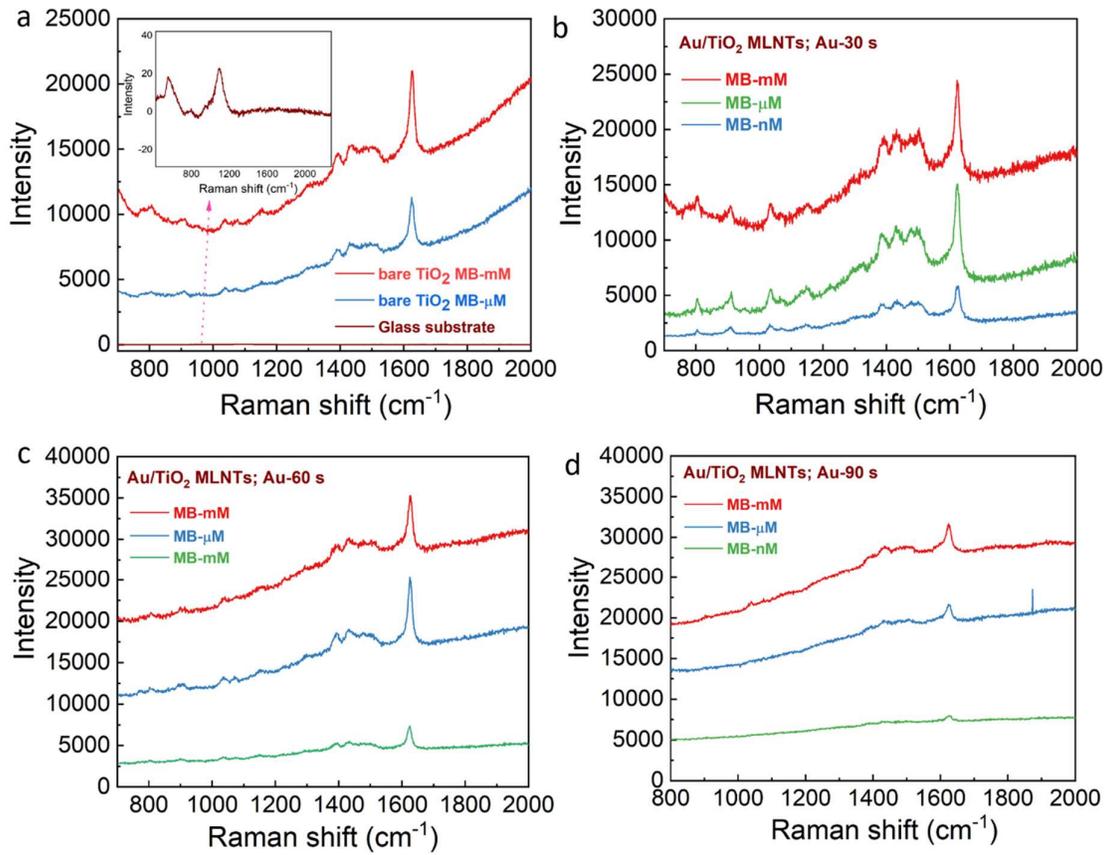

**Figure 4**: (a) SERS spectra of bare TiO$_2$ MLNTs and glass substrates modified with mM and µM concentrations of MB. (b) SERS spectra of Au/TiO$_2$ MLNTs modified with mM, µM, and nM concentrations of MB; Au-30 s. (c) SERS spectra of Au/TiO$_2$ MLNTs modified with mM, µM, and nM concentrations of MB; Au-60 s. (d) SERS spectra of Au/TiO$_2$ MLNTs modified with mM, µM, and nM concentrations of MB; Au-90 s.

Raman measurements were performed to examine the SERS sensitivity of the TiO$_2$ and Au/TiO$_2$ MLNTs using a probe analyte MB and are shown in Figure 4. Three molar concentrations (mM, µM, and nM) of MB were considered to analyze the SERS enhancement. Figure 4a demonstrates the SERS sensitivity of bare TiO$_2$ MLNTs in comparison with a non-SERS spectrum. Interestingly, bare TiO$_2$ MLNTs themselves showed significant enhancement in SERS activity for the detection of MB up to nM concentrations. Here, the SERS activity of bare TiO$_2$ MLNTs is mainly attributed to the enhanced light-matter interaction. Bare TiO$_2$

MLNTs displayed a broad absorption peak at ~ 535 nm, which is the photonic absorption edge of $TiO_2$ MLNTs. Here, it should be noted that the incident wavelength of the laser probe light is 532 nm. Therefore, during the SERS measurements, the incident laser light resonantly couples with the existing photonic absorption edge of $TiO_2$ MLNTs. Thus enhances the light-matter interaction significantly in the vicinity of the probe molecule adsorbed on $TiO_2$ MLNTs. As a result, the Raman signal of the probe molecule is enhanced several times, the same is displayed in Figure 4a. Therefore, we emphasize that the gradient refractive index profile associated with $TiO_2$ MLNTs played a crucial role in promoting the multiple reflections of light as well as generating the photonic stop bands for the enhanced light matter-coupling.

Similarly, the measured SERS spectra for different sputter deposition times of Au thin film (30-60 s) are displayed in Figure 4(b-d). As expected, the spectra reveal a further increase in the intensity of the MB Raman signal. For a better comparison, the baseline-corrected SERS spectra of bare and Au/$TiO_2$ MLNTs are presented in Figure 5a for mM concentration of MB. The intensity of peak appeared at 1632 $cm^{-1}$ following the trend of Au deposition times (30 s to 60 s), whereas the peak intensity is slightly decreased for 90 s deposition. Such an enhancement in the SERS signal is attributable to the combined role of photonic modes and plasmonic modes associated with the Au/$TiO_2$ MLNTs. Here, the synergy that exists between the LSPR modes coupled with the photonic modes of $TiO_2$ MLNTs plays a crucial role in multiplying the light-matter interaction.[32] The complementation among photonic and plasmonic modes associated with the metal and/or semiconductor nanostructures is also been reported for other applications.[14,32]

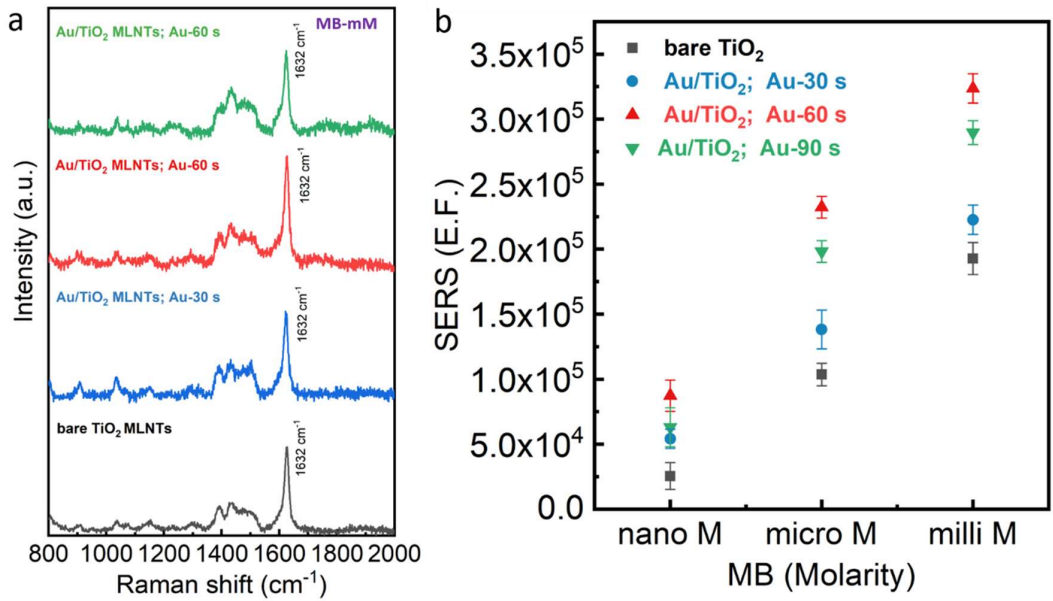

**Figure 5**: (a) Baseline corrected SERS spectra of TiO$_2$ and Au/TiO$_2$ MLNTs (MB-mM). (b) Enhancement factor trend of all prepared samples for all molar concentrations of MB.

Since the enhancement factor (E.F.) is a significant number in assessing the sensitivity of SERS substrates, here, we estimated the E.F. of our SERS substrates using the following formula.[33]

$$E.F. = \frac{I_{SERS}}{N_{SERS}} \times \frac{N_0}{I_0}$$

Where $I_{SERS}$ and $I_o$ are the intensities of the SERS signal observed at 1632 cm$^{-1}$ for bare TiO$_2$ MLNTs, Au/TiO$_2$ MLNTs, and a non-SERS substrate (glass slide). $N_{SERS}$ and $N_o$ are the number of probe molecules contributing to the SERS signal (TiO$_2$ or Au/TiO$_2$ MLNTs) and the non-SERS signal (glass substrate).

Since, we used similar MB deposition and SERS activity measurement (laser power, incident spot size, acquisition time, etc.) conditions, the number of probe molecules is assumed to be the same for all samples (TiO$_2$, Au/TiO$_2$ MLNTs, and glass) and distributed uniformly. The E.F. of all samples for all concentrations of MB is calculated by considering the SERS intensity of the baseline corrected predominant peak that appeared at 1632 cm$^{-1}$. For representation purposes, baseline corrected SERS spectra of $TiO_2$ and $Au/TiO_2$ MLNTs (Au, 30-90 s) for mM concentration of MB are shown in Figure 5a. The trend of E.F. presented in Figure 5b follows the order of Au layer deposition times 30 s and 60 s starting from the bare $TiO_2$ MLNTs. Whereas, for 90 s Au deposition, the SERS sensitivity slightly decreased for all molar concentrations of MB. The bare $TiO_2$ MLNTs showed an E.F. of ~ $2.5\pm1.1\times10^4$ for nM and mM concentration is ~ $1.7\pm1.2\times10^5$. This suggests the potential use of photonic semiconductor nanostructures as SERS substrates and the values are in line with the recent report on plasmon-free PBNs.[17] On the other hand, the $Au/TiO_2$ MLNTs (Au-60 s) has displayed the highest E.F. among all samples; for nM concentration, E.F. is $9.7\pm1.5\times10^4$, and for mM concentration it is $3.3\pm1.1\times10^5$. These values are approximately one order higher for nM, and two times higher for mM concentrations of MB in comparison to the bare $TiO_2$ MLNTs. Thus, the photonic and plasmonic bands collectively acted together to enhance the SERS signal of MB.

In conclusion, we demonstrate the potential use of semiconductor $TiO_2$ MLNTs as SERS substrates. Initially, $TiO_2$ MLNTs were fabricated using a simple single-step anodization method. Such nanotubes without any plasmonic support structure displayed a significant SERS sensitivity, which is attributed to the enhanced light-matter interaction facilitated by the layered structure of gradient refractive index morphology. In addition, to further improve the SERS sensitivity, we used a conformal coating of gold nanoparticle layer over $TiO_2$ MLNTs. The nanocomposite morphology ($Au/TiO_2$ MLNTs) exhibited improved SERS activity facilitated by the synergy that exists between photonic and plasmonic modes. Thus, the fabrication procedure and SERS substrates ($TiO_2$ MLNTs) presented in this work are simple and inexpensive and can be explored to probe analytes for physical, chemical, biological, and analytical applications.

## Acknowledgments

R.Y. acknowledges the European Commission for a Marie-Sklodowska Curie research fellowship grant (Project-PIERCAT) and the host Prof. Emiliano Cortes, Nanoinstitute Munich, University of Munich, Germany for fruitful comments and discussions.

## Author declarations

### Conflict of interest

The authors declare no conflicts of interest.

## Data availability

The data that supports the findings of this study are available with in the article.## References